# *Origami*-based Zygote structure enables pluripotent shape-transforming deployable structure

Yu-Ki Lee[1†], Yue Hao[2†], Zhonghua Xi[2], Woongbae Kim[3,4], Youngmin Park[1],

Kyu-Jin Cho[3,4], Jyh-Ming Lien[2*], In-Suk Choi[1*]

[1]Department of Materials Science and Engineering, Seoul National University, Seoul, 08826, Republic of Korea

[2]Department of Computer Science, George Mason University, Fairfax, VA, 22030, USA.

[3]Soft Robotics Research Center, Seoul National University, Seoul 08826, Republic of Korea.

[4]Department of Mechanical Engineering, Institute of Advanced Machines and Design, Institute of Engineering, Seoul National University, Seoul 08826, Republic of Korea.

†These authors contributed equally to this work

**Corresponding Author**

Jyh-Ming Lien (jmlien@cs.gmu.edu, Tel.: +01-703-993-9546, Fax: +01-703-993-1710)

In-Suk Choi (insukchoi@snu.ac.kr, Tel.: +82-2-880-1712, Fax: +82-2-883-8197)

**Abstract**

We propose an algorithmic framework of a pluripotent structure evolving from a simple compact structure into diverse complex 3-D structures for designing the shape transformable, reconfigurable, and deployable structures and robots. Our algorithmic approach suggests a way of transforming a compact structure consisting of uniform building blocks into a large, desired 3-D shape. Analogous to the pluripotent stem cells that can grow into a preprogrammed shape according to coded information, which we call DNA, compactly stacked panels named the zygote structure can evolve into arbitrary 3-D structures by programming their connection path. Our stacking algorithm obtains this coded sequence by inversely stacking the voxelized surface of the desired structure into a tree. Applying the connection path obtained by the stacking algorithm, the compactly stacked panels named the zygote structure can be deployed into diverse large 3-D structures. We conceptually demonstrated our pluripotent evolving structure by energy releasing commercial spring hinges and thermally actuated shape memory alloy (SMA) hinges, respectively. We also show that the proposed concept enables the fabrication of large structures in a significantly smaller workspace.

# 1. Introduction

Nature-inspired shape programmable structures have generated substantial interest in mathematics, applied physics, computer science and graphics, materials science, robotics, and biological engineering[1-7]. Diverse structures in nature are sometimes constructed by a combination of basic units under simple, rational algorithms[8-9]. For example, complex patterns on leaves are composed of a single geometry and its repetition called a fractal, e.g., veins on a leaf branch off in the form of a self-similar geometry[10]. The living organism also grows and evolves based on the combination of four nucleotides called DNA. This coded information, consisting of only four nucleotides, determines how small stem cells proliferate and grow into the preprogrammed shape (Figure 1a)[11-13]. In addition, proteins with complex 3-D structures are synthesized with a combination of only twenty amino acids. The preprogrammed information called codons in mRNA determines the sequence of the amino acids (Figure 1b)[14-16]. In other words, nature may consist of simple geometry and an algorithm determining their combination.

Analogous to these growing and evolving systems of nature, researchers have shown that the rational design of simple unit structures can realize shape-programmable structures[17, 18]. One of popular strategies is *origami* or *kirigami* (art of paper folding or cutting) structures. For example, Cho and colleagues suggest a fractal cut pattern that can program the shape of a *kirigami*-inspired deployable structure with only two cut motifs and their hierarchical arrangement[19]. Another strategy is developing rational algorithms that construct desired structure with building unit blocks. Luo and colleagues developed a computational framework named legalization, which constructs arbitrary 3-D structures by assembling basic building blocks[20], and Yu and colleagues also showed that a single-chain structure can transform into diverse 3-D structures by coiling and stacking layer-by-layer under computational guidance[21].

However, most previous studies (1) allowed the variation of unit blocks (e.g., building blocks have different sizes or shapes), (2) were based on the disassembly and reassembly of unit blocks, or (3) have limited deployability (e.g., the ability to expand its scale)[22-24]. In other words, it has not yet been proposed how we can develop a pluripotent shape-transformable structure, i.e., a compact structure consisting of uniform building blocks that can transform into diverse huge 3-D shapes.

In this study, we first propose an *origami* (art of paper folding)-inspired algorithmic approach for both shape-programmable structures (i.e, a structure that can change its shape into a desired form) and deployable structures (i.e., a structure that can change its size from a compactly small scale to a significantly large scale) consisting of uniform building blocks. Similar to the natural system, our shape-programming framework of pluripotent evolving structure, named a zygote structure, consists of fully connected thin, uniform building blocks. Controlling its connection path with a computationally designed coded sequence, it can be deployed into the preprogrammed structure.

## 2. Basic concept of Zygote structure

Figure 1c shows the basic concept of our pluripotent evolving structure called zygote structure. All the building blocks are uniform square panels in our system. Starting from a given compactly stacked panels, we connect them according to the preprogrammed connection path (i.e., the computationally designed coded sequence) that can expand it into a desired 3-D *origami* structure. A panel is physically connected with the upper or lower panel through one of its four sides by a rotational hinge. We constrained the rotation angles of the hinge to 90°, 180°, or 270° (Figure 1d). In our simulation for the shape transformation, the hinges are designed to be stretchable to avoid the folding errors caused by thick panels. After connection, the stacked panels become a fully connected single-chain structure (which will be called a "stripified panels" hereinafter) that can be unfolded and reconfigured into a 3-D *origami* structure. In this framework, the search space for finding the valid connection path is $4^N \times 3^N$ according to the number of panels N. Then, the key question is how we can find the coded sequence for a specific shape efficiently in this large search space. The computer science community has developed answers to this problem[21, 23]. Rather than searching for the stacked state in this large space, previous studies have focused on finding connection paths of panels tessellating the surface of 3-D structures before compacting them. Similar to these previous approaches, our stacking algorithm finds the path that can compactly stack 3-D structures to solve this inverse problem.

Figure 1e shows our stacking algorithm that transforms a 3-D structure into stacked panels in which all panels are connected in a single-strip structure. To achieve this, we first approximate a curved 3-D surface with square panels via voxelization[25]. Then, we represent the voxelized surface as a dual-graph G = <V, E>, where each node in V corresponds to a square panel and each edge in E corresponds to a rotating hinge connecting the two neighboring panels. This inverse problem can be solved by finding the Hamiltonian path P, i.e., a path that visits all of the vertices on the given G. In general, the Hamiltonian path problem (HPP) is NP-complete; however, Hamiltonian cycles must exist in the dual graph of the voxelized mesh as a 4-regular graph[26]. To solve this HPP, we used the traveling salesman problem (TSP) solver, which finds a Hamiltonian cycle, i.e., a path with identical start and end, with minimum cost[27]. Once a Hamiltonian cycle is found, we break the cycle at an arbitrary position. This inverse programming can be operated in almost linear time by the number of panels. Cut and folded along P, the tessellated 3-D surface can be transformed into stripified panels and stacked into a columnar pile. The stripified panels can be more compactly stacked with multiple piles, as shown in Figure 1f. The yellow boxes in Figure 1f show the conceptual configuration of the compaction process. Finally, we inversely use P as the connection path for shape programming, i.e., P for stacking can inversely guide the compactly stacked panels to deploy into the target 3-D structure. For example, Figure 1g shows the experimental result that compactly stacked square sheets connected by the coded sequence can deploy into a 3-D turtle structure that has a 144 times larger bounding box size. More importantly, we can deploy stacked panels into different 3-D structures just by re-arranging their connection path by finding alternative or different Hamiltonian paths for diverse 3-D surfaces as long as we can tessellate surface of a structure with the same number of panels.

In this aspect, we refer to given initial stacked panels as zygote structures to represent that this initial simple structure can deploy into diverse 3-D structures under the guidance of the coded sequence, analogous to the zygote cell in an organism that is divided and grows into the preprogrammed shape.

**3. Implementation of Tree-stacking algorithm into Zygote structure.**

In practical realization, stripified panels can result in significant inaccuracy for transforming the compactly stacked structure into the 3-D surface (or vice versa) because even small noises in the motion control of the folding hinges can be cumulated along the connection-path and cause large gaps or overlaps in the final structure[28]. This problem can be reduced by representing the zygote structure as a spanning tree T in the graph G rather than a single Hamiltonian path P. While all vertices in P are less than 2 degrees nodes (i.e., every panel can be physically connected with only two adjacent panels except the first and the last panel), the spanning tree T allows high-degree nodes (HDNs) with greater than 3 connections. Figure 2a shows the conceptual configuration of the zygote structure in which panels are connected with a tree-path T rather than a single Hamiltonian path P; panels in a pile are connected with a Hamiltonian path, and each pile is connected with hinges called bridges. The panels having bridges correspond to the HDNs in T.

Figure 2b details this tree stacking using an example of the plane sheet approximated with 12 × 12 quads (Note that our stacking algorithm is valid for not only closed 3-D structures but also unclosed structures such as a plane sheet). We first divide G with K subgraphs $G_s$. A Hamiltonian path P in $G_s$ can potentially stack the corresponding panels into a single pile, i.e., the number of partitions determines the number of piles in the zygote structure. Multiple P in $G_s$ increase the pluripotency of the zygote structure. One way to achieve this is to increase the number of edges, $|E(G_s)|$, within $G_s$ since the diversity of finding P also increases due to higher inner connectivity. We also aim that all the piles have the same heights, i.e., each subgraph has the same number of vertices. This increases the deployability (e.g., volume expandability) of the pluripotent evolving structure. This partitioning problem can be considered as a balanced graph partitioning problem[29], which divides a graph into K components having an equal or similar size while minimizing the total weight of the edges connecting different components. In our system, all the weights of edges are set to 1. In this case, the generalized Fiduccia-Mattheyses algorithm can easily minimize the number of edges in the cut[30]. However, the Fiduccia-Mattheyses algorithm starts from a random partition and optimizes it for the minimal cut length, so it does not guarantee an exactly equal number of vertices in K components and may result in unbalanced stacked panels (Figure 2c)[31]. To address this limitation, we repeat the Fiduccia-Mattheyses algorithm until it finds a balanced partition in which all components have the same number of nodes (Figure 2d). Although this approach is a quite simple and brute-force alternative, this method was sufficient for partitioning models with few hundreds to thousands of meshes into less than ten components.

After partitioning, we consider the placement of piles to maximize the connectivity between $G_s$. We set a hypergraph $G_{hyper}$ as a 2-D grid. Each node $v_{hyper}$ in $G_{hyper}$ corresponds to a subgraph $G_s^i$ and occupies a single grid cell of the 2-D grid, as shown in Figure 2e. When the number of edges in G between two subgraphs $\|E(G_s^i, G_s^j)\| > 0$, an edge $e_{hyper_{i,j}}$ connects $v_{hyper,i}$ and $v_{hyper,j}$ with the weight of the edge $W(hyper_{i,j}) = \|E(G_s^i, G_s^j)\|$. Then, we can find the optimal hypergraph $G'_{hyper}$ as

$$G'_{hyper} = \underset{e_{hyper} \in g}{\mathrm{argmax}} \sum W(e_{hyper}).$$

Then, each K component finds an arbitrary Hamiltonian path $P^i$ (Figure 2f). To construct a valid tree path, a Hamiltonian path $P^i$ should be adjusted to be connectable with its adjacent Hamiltonian path $P^j$ on both the 3-D structure and zygote structure, e.g., at least two adjacent nodes in each neighboring Hamiltonian path on the 3-D structure should also be adjacent in the stacked state. To enforce this requirement, we optimize $P^i$ locally in each pile $G_s^i$ to maximize connections between piles: (1) we select an arbitrary reference node $v_{hyper,r}$ in $G'_{hyper}$. Depending on their topological distance from $v_{hyper,r}$ in $G_{hyper}$, any child node $v_{hyper,i}$ is linked to its parent node $v_{hyper,j}$. (2) Let $v_x^i$ be the xth node in $P^i$; we place the stacked pile of $P^i$, which corresponds to $v_{hyper,i}$, next to $P^j$ with $v_1^i$ and $v_1^j$ being the bottom. Then, we recursively update the $P^i$ based on $P^j$ until there are some $v_x^i$ in $P^i$ that potentially be connected to node $v_x^j$ in $P^j$. This matching function is defined as

$$f_{match}(x) = \begin{cases} 1, & \text{if the edge } v_x^i, v_x^i \text{ exists in } G \\ 0, & \text{otherwise} \end{cases}$$

which checks every pair of nodes $v_x^i$ and $v_x^j$ in $P^i$ and $P^j$ that are on the same height x in the stacked configuration. (3) We update $P^i$ using

$$P^i = \operatorname*{argmax}_{P^i \text{ in } G_s^i} \sum_{x=1}^{\min(H(P^i), (H(P^j))} f_{match}(x) \cdot f_{geo\ match}(x)$$

where $f_{geo\ match}(x)$ is the geometric matching function (Figure 2g shows that $P^B$ is updated to topologically match $P^A$. All $P^i$ are recursively updated in this way). If there is already a hinge connecting $v_x^j$ and $v_{x+1}^j$ where the potential hinge between $v_x^i$ and $v_x^j$ should be in the stacked configuration, they cannot be linked since connecting them will result in nonmanifold geometries. If $f_{match}(x)$ violates this constraint, $f_{geo\ match}(x) = 0$; otherwise, $f_{geo\ match}(x) = 1$. (4) To ensure this geometric matching constraint, we locally cut the $P^i$ and change the orientation of $v_x^i$. For this process, we record all the xs that satisfy $f_{match}(x)=1$ and put in a sorted vector $\vec{x}$ and ensure $x_i < x_{i+1}$. If $|\vec{x}| = 1$, which means that there is only one matching component, we can simply rotate the entire pile along with $v_x^i$. However, if $|\vec{x}| > 1$, $v_x^i$ has to be detached from $v_{x-1}^i$ or $v_{x+1}^i$, i.e., we break $P^i$. We denote such a broken edge as $e_{break}^i$. Then, $v_x^i$ and $v_x^j$ can be connected by an edge $e_{match}^i$ connecting the neighboring piles. (Figure 2h) We call such $e_{match}^i$ as a "bridge". In our pluripotent evolving structure, these HDNs having three neighbors, including the bridge, correspond to the HDPs having three hinges (e.g., a hinge corresponding to the bridge connects panels in two adjacent piles, while the others connect the upper or lower panels). Finally, we obtain a zygote structure in which all panels are connected in the tree path as

$$T_s = \cup P^i + \cup E_{match}^i - \cup E_{break}^i \text{ for all i in}\{1, k\},$$

where $E_{match}^i$ and $E_{break}^i$ are the sets of all matching and broken edges in pile $P^i$. This tree-stacking algorithm effectively reduces the large gaps or overlaps in the final structure by preventing the accumulation of folding errors[28].

The tree-path $T_s$ generated by the tree-stacking process can inversely guide the zygote structures having the same configuration of panels into an arbitrary target 3-D structure. For example, Figure 2i shows that a voxelized 3-D fish model approximated with 144 square meshes is stacked into a zygote structure with 4-piles, which is the same as the zygote structure corresponding to the plane model approximated with 12 × 12 meshes (Figure 2j). This implies that we can develop diverse structures from a single zygote structure by controlling only the connection path and folding angles of hinges guided by the coded sequences, e.g., the $T_s$ can be used as a coded sequence for our pluripotent evolving structure. Since our main objective is transformation from a 3-D meshed structure into the stacked piles, it should be noted that the panels may overlap, if we flatten the zygote structure to a 2-D plane along the tree-path in contrast with the polyhedron unfolding algorithm that flattens a 3-D meshed structure on a 2-D plane without self-overlapping[32, 33]. If it is desired to create the flattened configuration and net a planar structure, we can require a 2-D grid such that each node occupies a single grid cell, and all nodes are connected along $T_s$. For instance, in case of the $T_s$ corresponding to the 3-D fish approximated with 144 panels, there are grid cells that are occupied by more than two nodes, as shown in Figure 2k (e.g., more than two panels can overlap in the flattened configuration). Even a single Hamiltonian path P can generate its self-overlapped planar figure, as shown in Figure 2l.

## 4. Demonstration of Algorithmic Zygote structure

Figure 3 shows examples of our zygote structure. According to the coded sequence generated by the tree-stacking algorithm, a zygote structure having 4,000 panels with 4 piles can evolve (e.g., deploy) into a chicken, a vehicle, and a fish model (Figure 3a). Besides pluripotency, scale expandability is another advantage of the pluripotent evolving structure. It is obvious that the size of the evolved structure is related to the thickness of each panel. To quantitatively compare the volume compaction/expansion ratio, we used package volume, which is widely used in the shipping industry[34]. Package volume measures the cuboid box size to compactly pack an object. In other words, the package volume expansion ratio (referred to as "VER" hereinafter), defined as $\frac{W_e \times L_e \times H_e}{W_z \times L_z \times H_z}$, measures the ratio of bounding box volume (Length × Width × Height) between the zygote and evolved 3-D structures. When the length of square panel L is 100 times the thickness t (e.g., L = 100 × t), the VER is 791, 431, and 1,077 for each chicken, vehicle, and fish model, respectively.

The number of panels (e.g., resolution) also affects VER. In general, the feasibility of tree stacking for high-resolution models is determined by the number of partitions rather than the number of panels. More partitions are much more expensive to compute, as each pile must have at least one HDN, or the entire algorithm must restart. In addition, the equal size partitioning process is also expensive for more partitions since we must repeat the Fiduccia-Mattheyses algorithm. When we increased the number of piles to more than 6 for the 4,000 panels case, the algorithm fails to find the solution. Meanwhile, our implementation using the state-of-the-art TSP solver Concorde[27] could find $P_s$ in each partition efficiently so that our pluripotent evolving structure can have high-resolution 3-D models if we restrict the number of partitions.

For experimental demonstration, we generate a pluripotent evolving structure with 92 panels (Figure 3b). A zygote structure with 2 piles can evolve into a vehicle having 13.7 VER or a fish having 15.6 VER when the length of square panel L is 20 times the thickness t (e.g., L = 20 × t). We prepared 92 plastic panels (20 cm × 20 cm × 1 cm size), triangular magnets for 90° and 270° rotating hinges, and cuboid magnets for 180° rotating hinges. We attached the magnets on the side faces of the plastic panels according to the coded sequence of the 3-D vehicle model. Then, we could easily transform it into the 3-D vehicle guided by the coded sequence generated by the tree-stacking algorithm. Although only 4 panels meet the ground (wheels of the vehicle) and there are no internal supports, the final vehicle structure stably maintains its shape without collapse. The 3-D vehicle can be folded back to the initial zygote state, and we can transform it again into the 3-D fish model just by rearranging the magnets (e.g., simply detaching the magnets and reattaching them) according to its coded sequence (Figure 3c).

## 5. Practical implications of Zygote structure

We conceptually show that our zygote structure can be used as the fabrication platform for overcoming the limited working space. Fabrication of a large structure may require a large area larger than the final product, e.g., making a large structure difficult and sometimes impossible without assembly due to limited workspace area or volume. Deployable structures that can overcome these limitations are suggested in diverse research fields involving architecture and robotics; however, most of them consist of bars and construct open scaffold structures[35-37]. In contrast to previous deployable structures, our pluripotent evolving structure can represent the external shape of the desired structures with thin panels. Figure 4a-d illustrates that a 3-D printed zygote structure having 92 panels with 2 piles can evolve into objects that are significantly larger than the printable size of the 3-D printer. Our 3-D printed panel has a size of 9 cm × 9 cm × 0.48 cm and has holes in its rounded side so that it can embed a cylindrical magnet with a diameter of 2 mm and length of 3 cm, as shown in Figure 4b. These cylindrical magnets are used as universal hinges for all 90°, 180°, or 270° rotations, which enables shape reconfiguration without rearranging the hinges (e.g., without detaching and reattaching the hinges in the zygote state). In addition, our panel design enables inserting magnets even inside of the printed zygote structure. Since all holes aligned along the four sides of each panel are exposed in the stacked configuration, we can insert magnets as the compactly printed configuration. According to the algorithmic result, we printed a zygote structure with 92 panels and 2 piles (Figure 4a). For printing, we introduced small gaps between panels (0.2 mm) to prevent their adhesion. The final printed zygote structure has a size of 23 cm × 18.2 cm × 9 cm. Then, we inserted the cylindrical magnets in all of the holes of the panels (total 368 magnets). As shown in Figure 4c, it could evolve (e.g., deploy) into an 81 cm × 72 cm × 18 cm airplane model with 31.3 VER by folding panels along its corresponding coded sequence. It could be folded back into the zygote state without an additional detaching process and could evolve again into a 63 cm × 27 cm × 27 cm size fish model (Figure 4d). Both 3-D structures have larger dimensions than the printable size (23 cm × 19 cm × 20 cm, marked in red in Figure 4c-d). We acknowledge that the pluripotent evolving structure with higher resolution may practically result in unstable or floppy structures because it deploys into hollow 3-D structures; however, this issue can practically be mitigated by adding additional hinges connecting the cut edges only at the target state or adding internal support. For robust final structures, it may potentially be integrated with internal folding struts inside the unit panels or recent hydraulic- or motor-based moving hinges[38-40].

Our demonstration with a 3-D printer further implies that our zygote structures can be incorporated with recent 4-D printing techniques[41-43] or *origami* robots[44-46]. The zygote structure can easily incorporate these self-transformation systems because thin panels as building blocks have advantages in terms of light weight and integration with novel stimuli-responsive materials than volumetric building blocks[17, 45-49]. To integrate the *origami* robot techniques in our zygote structure, we introduce energy-releasing spring hinges and thermally actuated shape memory alloy (SMA) hinges in our pluripotent evolving structure. We are the first to demonstrate a shape-reconfigurable structure that can transform into multiple targets starting from a compact structure consisting of a thin, uniform panel. Figure 4e shows a simple zygote structure having 14 panels with 2 piles that can be deployed into an L- or an I-shaped 3-D structure. We experimentally demonstrate the deployable zygote structure by assembling 3-D printed 14 panels with commercial spring hinges; their rotation angles were controlled to be 90°, 180°, or 270° by welding steel bars on the spring hinges (Figure 4f). The spring hinges can be inserted into one of four sides of the 3-D printed panels. First, we connected stacked panels with the spring hinges according to the coded sequence for the I-shape structure. Simply thrown into the air, it rapidly deploys into the final 3-D structure, and the unfolded structure maintains its shape without additional supports. Then, we could transform it to the zygote structure easily by folding back guided by the hinges and detaching them and reconfiguring it into the L-shape structure in the same way (Figure 4g). This demonstration using spring hinges enables fast and robust shape transformation, while recent stimuli-responsive self-folding systems experience delays caused by phase transformation or alignment of materials[17].

We further demonstrate that our pluripotent evolving structure can be combined with recent self-folding systems based on novel stimuli-responsive materials. Most of these systems are composed of functionalized or layered thin sheet-like structures[47-49]. In other words, whether incorporating thin sheet-like hinges is possible is the key criterion to test the potential feasibility of introducing stimuli-responsive hinges. To demonstrate this, we simply introduce shape memory alloy (SMA) sheets into our pluripotent evolving structure. We prepared 5 mm × 10 mm thin SMA sheets. They are folded to 90° and fixed, annealed at 450 °C for an hour, and quenched in water. This process preprograms the shape of SMA sheets. Overlapping the programmed SMA by folding it in half with either 90° or 270°, we could prepare SMA hinges that can rotate and stop either at 90° or at 270°. Hinges for 180° rotation are prepared in the same manner (Figure 4h). Then, we attached them to each of two zygote structures consisting of 14 pieces of papers guided by the coded sequence for each I- or L-shaped structure. Figure 4i shows that the pluripotent evolving structure can deploy into the two 3-D structures when we apply heat by a commercial heat gun. Stimuli-responsive materials have enabled diverse shape-transformable structures, but its integration with a computationally guided design system as in our study have been rarely reported. Although we simply demonstrate the pluripotent evolving structure with spring hinges or SMA hinges, we believe that recent *origami* robots that embed circuit boards, sensors, or batteries can be applied for realizing a functionalized shape-programmable *origami* robot[44-46].

## 6. Summary

We proposed the concept of the pluripotent evolving structure for designing a shape transformable, reconfigurable, and deployable structure. Computational inverse design using our tree-stacking algorithm enables finding the connection paths that inversely guide compactly stacked uniform panels called the zygote structure into an arbitrary 3-D structure. We demonstrated that our framework provides a shape-programmable structure with both high pluripotency and deployability. For example, we showed that this inverse design approach could transform a compact zygote structure having 4,000 panels into a chicken, a vehicle, and a fish model having 791-, 431-, and 1077-times package volume sizes than their initial state. Thin building blocks in our zygote structure not only lead to extremely high deployability (e.g., volume expandability) but also potentially bring better feasibility for self-transformation combined with diverse self-folding mechanisms based on thin-film materials. We experimentally demonstrated the self-transformation of our pluripotent evolving structure with commercial spring hinges and SMA sheets. This further implies that our self-transformation system can be further incorporated with recent 4-D printing systems. In addition, it has versatile application rather than previous shape-programmable structures since conventional functionalized thin materials or fabrication processes for thin film materials can be easily incorporated into the thin, uniform unit panels of our pluripotent evolving structure. One potential application is the design of wearable devices and robots, in which our pluripotent evolving structure provides tight conformability to the underlying shapes or human body. Recent autonomous origami robots embedding circuit boards or microcontroller can be applied to achieve it[44-46]. In these aspects, our concept for pluripotent evolving structures not only contributes to the development of rational algorithms for shape-programmable structures in computer science and graphics, materials engineering, and architectures but also provides new insights into the development of portable, deployable, or 3-D shaped devices and robots in engineering fields involving materials engineering, electronic engineering, aerospace engineering, and robotics.

**Acknowledgments**

This work was supported by the National Research Foundation of Korea (NRF) (grant numbers 2019R1A2C2003430, 2018M3A7B4089670, 2016R1A5A1938472), the Creative-Pioneering Researchers Program through Seoul National University, and LG Display under the LGD–Seoul National University Incubation Program.

**Author Contributions**

I.-S. Choi conceived the concept. Y. Hao, Z. Xi, and J.-M. Lien developed the stacking algorithms. Y.-K. Lee and Y. Hao generated the 3-D models used in this paper and manipulated and implemented the algorithm. Y.-K. Lee and Y. Park carried out experimental demonstrations. W. Kim and K.-J. Cho discussed and commented on the experimental demonstration. J.-M. Lien and I.-S. Choi supervised the project. Y.-K. Lee, Y. Hao, Z. Xi, J.-M. Lien, and I.-S. Choi wrote the paper. Competing interests: The authors declare no competing interest.

**Notes**

The authors declare no competing financial interest.

**Materials and Methods**

We implemented the proposed stacking algorithm in C++. All data are collected on a MacBook Pro with a 2.5 GHz Intel Core i7 CPU with 16 GB memory running macOS 10.12. For the 3-D printed panels, we used commercial PLA (Polylactic acid) filaments. For the experimental demonstration with the shape memory alloy sheets, we used 0.125 mm thick nitinol foil (purchased from AVENTION Co., Ltd.)

**Data and materials availability**

All data needed to evaluate the conclusions in the paper are presented in the paper. Additional data related to this study are available from the corresponding author upon request.

**Code availability**

Computer codes used in this study are available from the corresponding author on request.

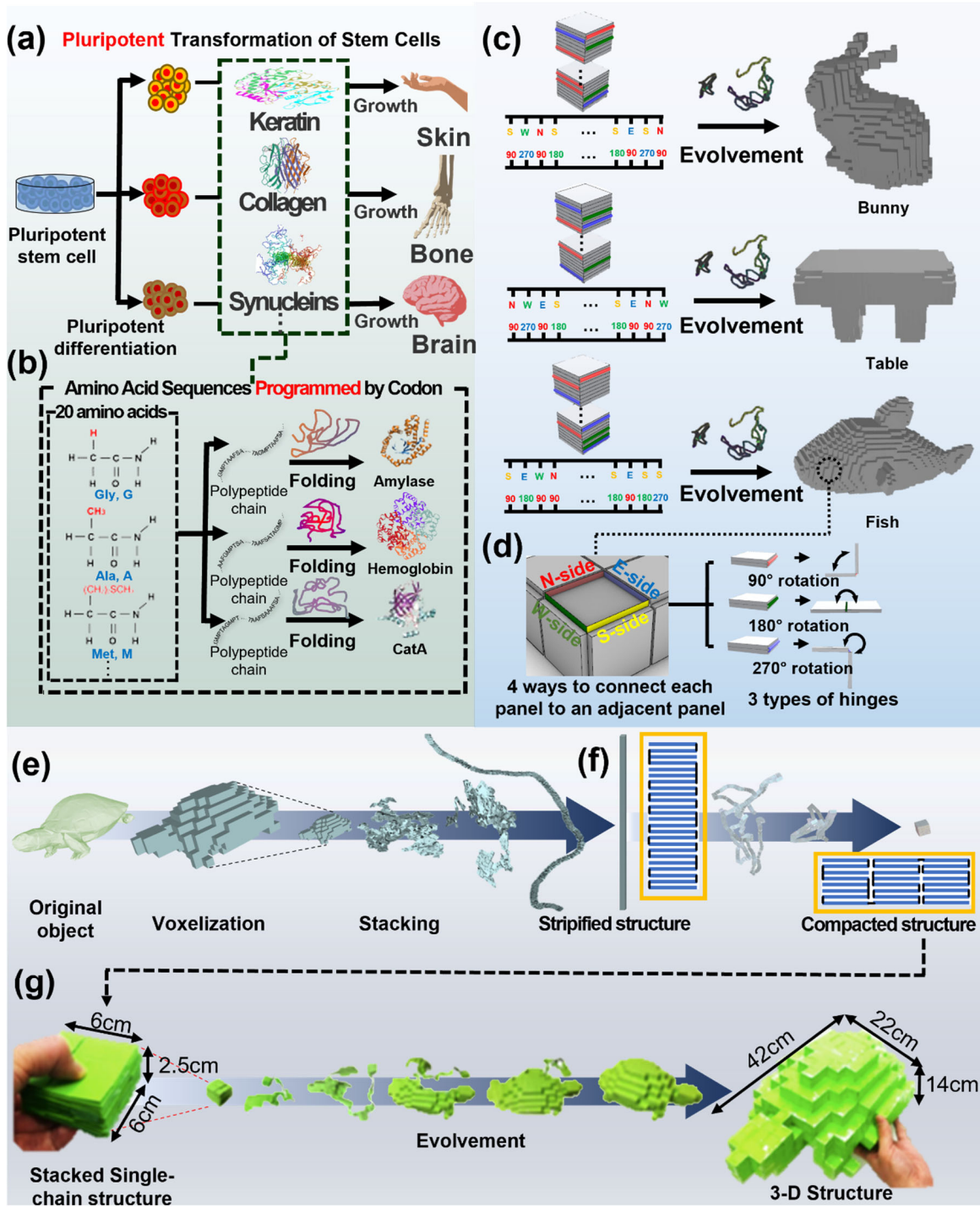

**Figure 1. Concept of the pluripotent evolving structure. a.** Tiny pluripotent stem cells differentiate into preprogrammed diverse shapes, e.g., they differentiate and proliferate into diverse proteins and organs. The DNA consisting of only four nucleotides guides them to the final shapes. **b.** Codons in mRNA can also program the shape of proteins, which is the sequential connection of only twenty amino acids. A polypeptide chain, which is a single chain in which twenty amino acids are connected in order of the preprogrammed sequence in the codon, transforms into its corresponding 3-D structure called a protein. **c.** Analogous to these shape programming systems in nature, our pluripotent evolving structure starts from stacked

panels named the zygote structure, and it deploys into final structures guided by coded information. **d.** Our stacking algorithm inversely finds the connection path and folding angles of each panel in the zygote structure to transform it into diverse 3-D shapes in a large space increasing as $4^N \times 3^N$ according to the number of panels N. **e.** The stacking algorithm approximates a target 3-D surface with square panels first and compactly stacks them with a single-strip structure by finding a Hamiltonian path on a dual graph on the target surface. **f.** The single-chain structure can be further compacted by dividing the stripified panels into K components. **g.** By connecting paper panels according to this connection path, we could prepare a compacted structure and deploy it into a 3-D turtle model, which has an almost 144 times larger scale.

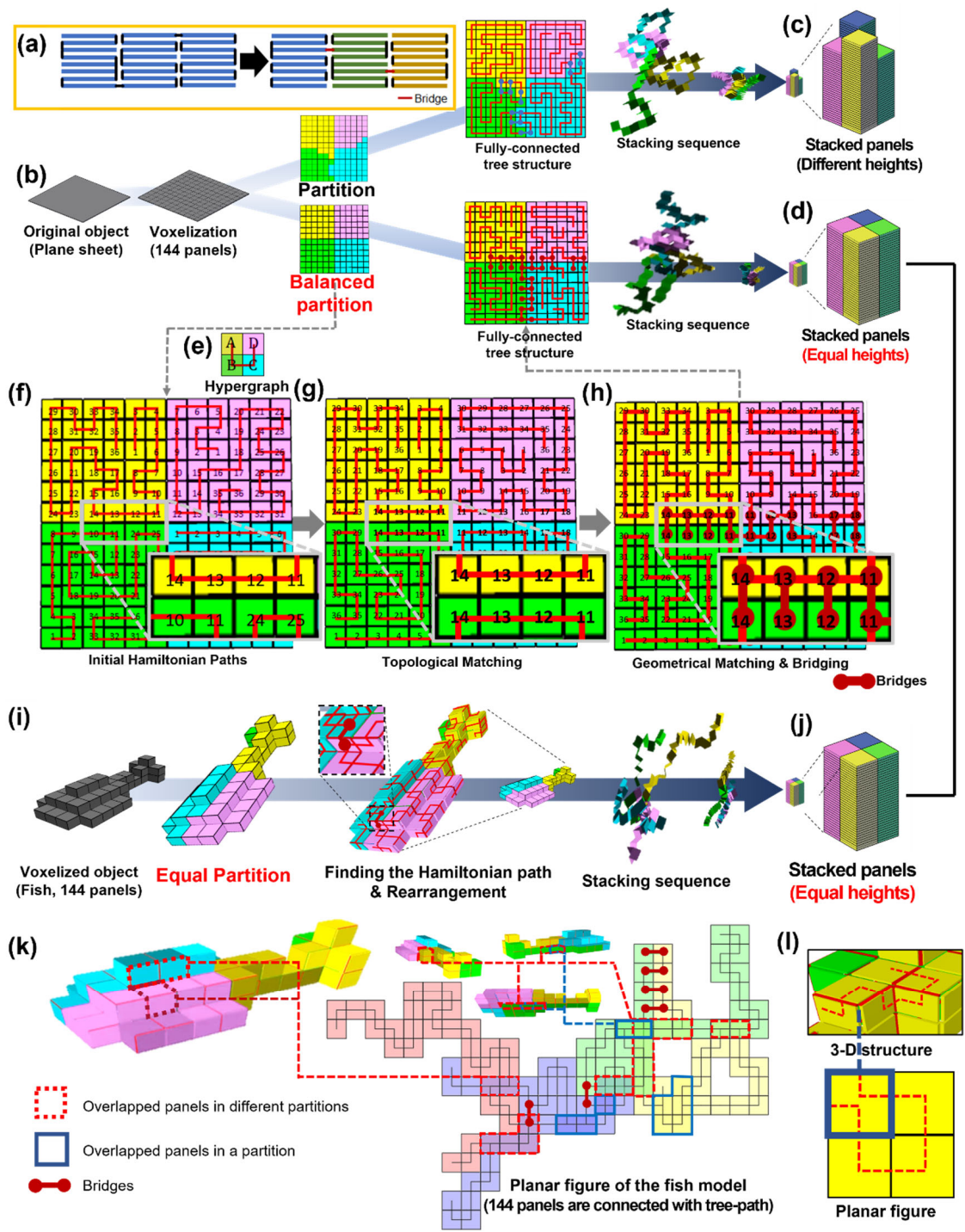

**Figure 2. The tree stacking algorithm with balanced partition. a.** Rather than stripified panels corresponding to a Hamiltonian path, the tree-path structure reduces cumulative folding errors for nonzero thick panels and provides better load-bearing capabilities for practical application. **b.** For a given plane sheet structure approximated with 12 × 12 square meshes, our tree-stacking algorithm partitions the graph on it using the Fiduccia-Mattheyses algorithm. **c.** However, the Fiduccia-Mattheyses algorithm does not guarantee an equal number of panels in each partition and results in nonuniform heights of piles in the zygote structure. **d.** To obtain

the zygote structure in which all piles have the same heights for maximized pluripotency and deployability (e.g., volume expandability), we simply repeated this partition process until it generated a uniformly balanced partition. **e.** Each partition is stacked into a single pile, and each pile is connected with one of its adjacent piles to construct a fully connected tree-path. For this process, we set a hypergraph that determines the placement and connection between piles on a 2-D grid. **f.** Then, we find a Hamiltonian path on each partition. The number on panels shows the sequence of each panel in the Hamiltonian path (corresponding to the height in the stacked configuration), i.e., the numbers on the panels indicate the sequence of vertices $v_x$, e.g., a panel having number $X$ corresponds to the xth node in $P^i$. **g.** The Hamiltonian paths are adjusted to make a bridge that can connect two neighboring piles in both unfolded 3-D and stacked configuration (Hamiltonian path on partition B is adjusted and the 11th to 14th panels in A and B partitions become neighbors in both 3-D and stacked configuration). They are connectable with hinges called bridges. **h.** Finally, the geometric matching process checks potential panels having double hinges on a single side caused by the bridge and locally adjusts it to a valid connection path. **i.** When we apply this process for the 3-D fish model having the same number of panels (144 panels), **j.** it generates the same zygote structure, which implies that a zygote structure can deploy in both the plane sheet structure or the fish structure just by controlling their connection path. In our tree-stacking algorithm, we allow the overlapping of its unfolded figure. Both overlapped panels **k.** between different partitions or **l.** in a partition can be seen in the flattened configuration of the tree-path.

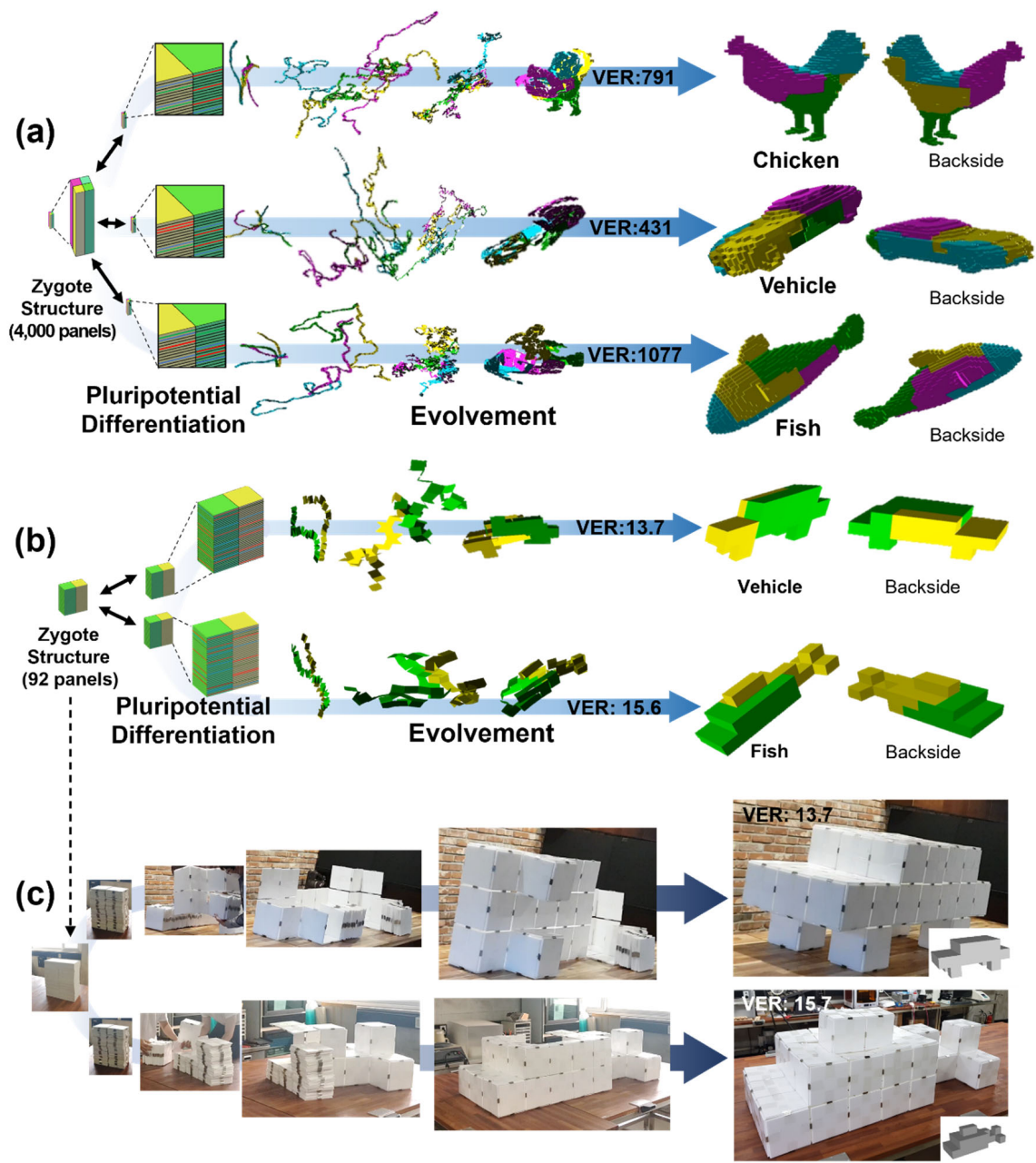

**Figure 3. Algorithmic results of the pluripotent evolving structure and its demonstration with magnets. a.** The coded sequences generated by the tree-stacking algorithm can guide the zygote structure (e.g., compactly stacked panels) into diverse 3-D structures, such as a chicken, a vehicle, and a fish. When the length of each panel is 100 times the thickness (L = 100 t), the package volume expansion ratio (VER) is 791, 437, and 1077 for each model (see Fig. S7 for a further explanation of VER). **b.** We simply demonstrated our pluripotent evolving structure with 92 panels. A zygote structure can evolve (e.g., deploy) into a vehicle and fish model, each having 13.7 and 15.6 of VER (L = 20 t) according to the coded sequences. **c.** We could simply demonstrate it with plastic panels by attaching triangular or rectangular magnets on the side of the panels. The final structures were robust and reliable without collapse, especially even for the vehicle structure in which only 4 panels were placed on the ground.

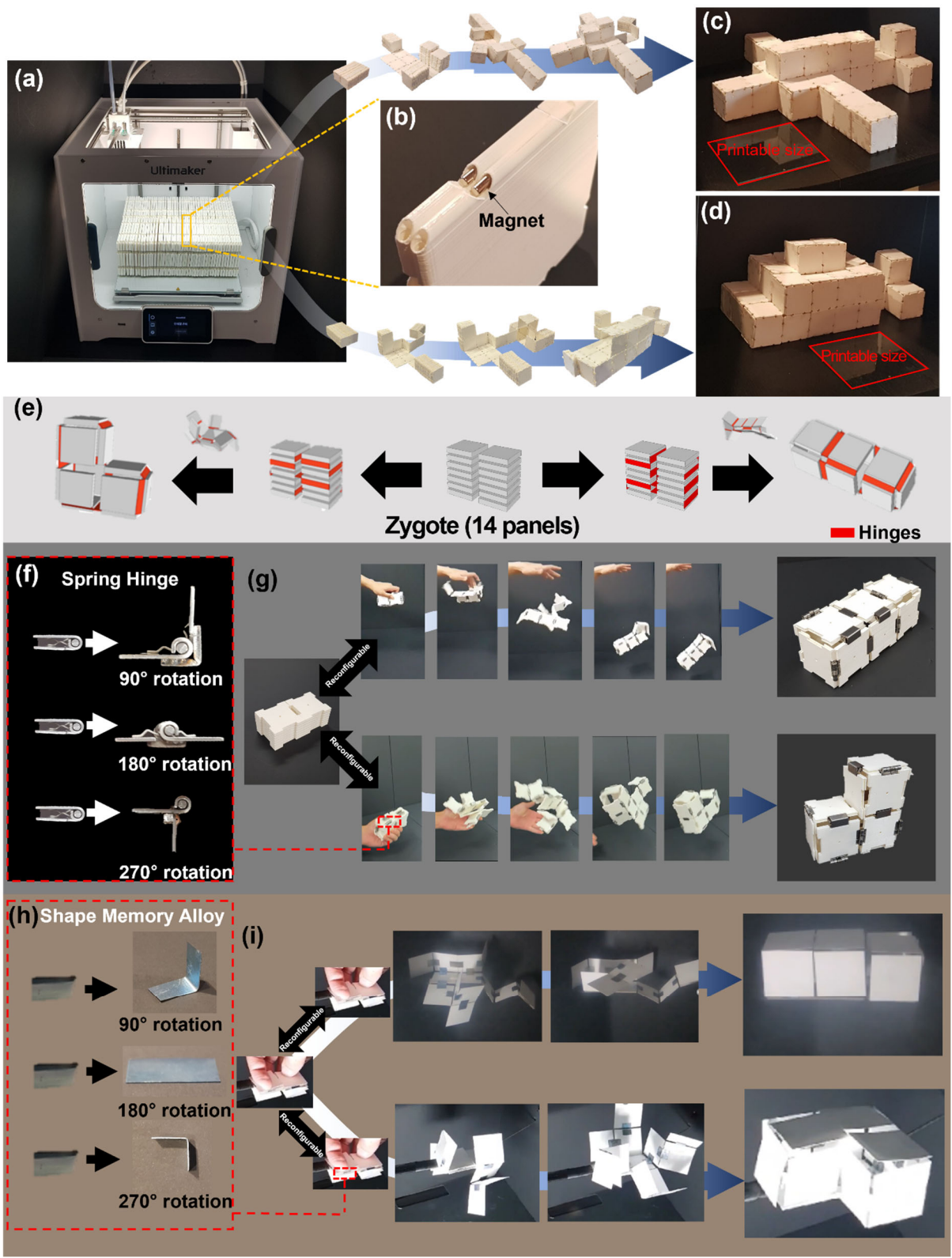

**Figure 4. Self-transformable pluripotent evolving structure and a conceptual demonstration of overcoming the limited fabrication space using the pluripotent evolving structure. a.** Our pluripotent evolving structure enables overcoming the limited workspace

since it can be fabricated in a compacted zygote structure and evolve (e.g., deploy) into huge structures. For example, we printed a zygote structure with 92 panels 23 cm × 18 cm × 9 cm in size. **b.** Each panel (9 cm × 9 cm × 0.48 cm size) can embed cylindrical magnets in its holes in its rounded sides (See Fig. S9). Without an additional reconfiguration process (e.g., detaching and reattaching hinges), **c.** the printed pluripotent evolving structure evolves into an airplane model (81 cm × 72 cm × 18 cm, VER = 31.3), is folded back into the zygote structure, **d.** and then evolves again into a fish model (63 cm × 27 cm × 27 cm, VER = 15.6). Both 3-D structures have more than 12 times or 5 times larger package volume sizes than the printable size (marked red) of the commercial 3-D print (23 cm × 19 cm × 20 cm). **e.** We also demonstrate self-deployable pluripotent evolving structure with a zygote structure having 14 panels with two piles that can evolve into L- or I-shaped 3-D structures without self-collision. **f.** We prepared three types of mechanical hinges that rotate 90°, 180°, or 270° by welding a steel bar on commercial spring hinges. **g.** Attached the spring hinges according to the coded sequences, the zygote structure rapidly evolves into the L-shape structure when we simply throw it into the air. It is also possible to fold back into the zygote state and reconfigure it by rearranging hinges. Then, it transforms it into an I-shaped structure in the same manner. **h.** We also showed that stimuli-responsive materials can be introduced in our pluripotent evolving structure. The SMA sheets are preprogrammed to be folded at 90°, 180°, or 270°. **i.** By attaching them to the stacked papers according to the coded sequences, we could realize a stimuli-responsive self-evolving structure.